\documentclass[fleqn,12pt,twoside]{article}
\usepackage{espcrc1}

\usepackage{graphicx}
\usepackage[figuresright]{rotating}

\newcommand{\AmS}{{\protect\the\textfont2
  A\kern-.1667em\lower.5ex\hbox{M}\kern-.125emS}}

\title{Nuclear structure calculations with a separable approximation
       for Skyrme interactions}

\author{A.P. Severyukhin\address[BLTP]{Bogoliubov Laboratory of Theoretical Physics,
        Joint Institute for Nuclear Research, \\
        141980 Dubna, Moscow Region, Russia},%
         V.V. Voronov\addressmark[BLTP],
         Ch. Stoyanov\address{Institute for Nuclear Research and Nuclear Energy,
         boul. Tzarigradsko Chaussee 72, 1784 Sofia, Bulgaria}
        and
        Nguyen Van Giai\address{Institut de Physique Nucl\'eaire,
        Universit\'e Paris-Sud, \\F-91406 Orsay Cedex, France}}

\begin{document}

% typeset front matter
\maketitle
\begin{abstract}
A finite rank separable approximation for the quasiparticle RPA
calculations with Skyrme interactions that was proposed in
our previous work is extended to take into account the coupling between one-
and two-phonon terms in the wave functions of excited states.
It is shown that characteristics  calculated within the suggested
approach are in a good agreement with available experimental data.
\end{abstract}

\section{INTRODUCTION}

Many properties of the collective nuclear excitations can be described
within the random phase approximation (RPA) \cite{R70,BM75,Schuck,solo}.
The most consistent models employ the Gogny's \cite{gogny} or
Skyrme-type \cite{vau72} effective interactions which
can describe the ground states in the
framework of the Hartree-Fock (HF) approximation and the excited states
within the RPA.
Such models are quite successful
for predicting nuclear states properties \cite{floc78,doba96,colo1,KG00,colo2}.

Due to the anharmonicity of vibrations there is a
coupling between one-phonon and more complex states \cite{BM75,solo}
and the
complexity of calculations beyond standard RPA increases rapidly
with the size of the configuration space, so one has to work within limited
spaces.
Using simple separable forces
one can perform calculations of nuclear characteristics in very large configuration
spaces since there is no need to diagonalize matrices whose dimensions grow
with the size of configuration space.
For example, the well-known quasiparticle-phonon model (QPM) \cite{solo}
can do very detailed predictions
for nuclei away from closed shells\cite{gsv}.

The possibility for such a simplification was the motivation for proposing
in our previous work \cite {gsv98} a finite rank approximation
for the particle--hole (p-h) interaction resulting from
Skyrme-type forces.
Thus, the self-consistent mean field can be calculated
in the standard way with the original Skyrme interaction whereas the RPA
solutions would be obtained with the finite rank approximation to the p-h~matrix
elements.
It was found that the finite rank approximation reproduces reasonably well
the dipole and quadrupole strength distributions in Ar isotopes \cite {gsv98}.

Recently, we extended the finite rank approximation for p-h interactions
of Skyrme type to take into account pairing \cite{ssvg02}.
In this paper we generalize our approach to take into account a coupling between the one-
and two-phonon components of wave functions.
As an application we present results of our first
calculations for the quadrupole and octupole states
in $^{112}$Sn.

\section{METHOD OF CALCULATIONS}

We start from the effective Skyrme interaction\cite{vau72}
and  use the notation of Ref.\cite{sg81} containing explicit
density dependence and all spin-exchange terms.
The single-particle spectrum is calculated within the HF method.
The continuous part of the single-particle spectrum is
discretized by diagonalizing the HF hamiltonian on
the harmonic oscillator basis\cite{BG77}.
The p-h residual interaction $\tilde V_{res}$
corresponding to the Skyrme force and including both direct and exchange
terms can be obtained as the second derivative of the energy density
functional with respect to the density\cite{ber75}.
Following our previous paper\cite{gsv98} we  simplify $\tilde V_{res}$ by
approximating it by its Landau-Migdal form.
For Skyrme interactions all Landau parameters $F_l, G_l, F^{'}_l, G^{'}_l$
with $l > 1$ are zero. Here, we keep only the $l=0$ terms
in $V_{res}$ and in the coordinate
representation one can write it in the following form:
\begin{eqnarray}
V_{res}({\bf r}_1,{\bf r}_2)=N_0^{-1}\left[ F_0(r_1)+G_0(r_1)
{\bf \sigma}_1{\bf \sigma}_2+(F_0^{^{\prime
}}(r_1)+G_0^{^{\prime }}(r_1){\bf \sigma }_1{\bf \sigma}_2){\bf
\tau }_1{\bf \tau }_2\right] \delta ({\bf r}_1-{\bf r
}_2)  \label{eq2}
\end{eqnarray}

where ${\bf \sigma}_i$ and ${\bf \tau}_i$ are the
spin and isospin operators,
and $N_0 = 2k_Fm^{*}/\pi^2\hbar^2$ with $k_F$ and $m^{*}$ standing for the
Fermi momentum and nucleon effective mass.
The expressions for
$F_0, G_0, F^{'}_0, G^{'}_0$ in terms of the Skyrme force parameters can
be found in Ref.\cite{sg81}. Because of
the density dependence of the interaction the Landau parameters of
Eq.(\ref{eq2}) are functions of the coordinate ${\bf r}$.

The p-h residual interaction can be presented as a sum of
$N$ separable terms.
Let us explain this procedure for making the finite rank approximation
by examining only the contribution of the term  $F_0$.
In what follows we use the second quantized representation
and $V_{res}$ can be written as:

\begin{eqnarray}
\hat V_{res} & = & \frac 12\sum_{1234}V_{1234}:a_1^{+}a_2^{+}a_4 a_3:
\end{eqnarray}

where $a^+_1$ ($a_1$) is the particle creation (annihilation) operator
and $1$ denotes the quantum numbers $(n_1l_1j_1m_1)$,

\begin{eqnarray}
V_{1234} = \int \phi^*_1({\bf r}_1)\phi^*_2({\bf r}_2)
V_{res}({\bf r}_1,{\bf r}_2)\phi_3({\bf r}_1)
\phi_4({\bf r}_2) {\bf dr}_1{\bf dr}_2 ,
\end{eqnarray}
\begin{eqnarray}
V_{1234}=\sum_{JM}\hat J^{-2}
\langle j_1||Y_J||j_3\rangle \langle
j_2||Y_J||j_4\rangle I(j_1j_2j_3j_4)
\times\\\nonumber
(-)^{J+j_3+j_4-M-m_3-m_4}\langle j_1m_1j_3-m_3 \mid J-M\rangle
\langle j_2m_2j_4-m_4\mid JM\rangle.
\end{eqnarray}

In the above equation, $\langle j_1 \vert\vert Y_{J} \vert \vert j_3 \rangle$ is
the reduced matrix element of the spherical harmonics $Y_{J \mu}$,
$\hat J = \sqrt {2J+1}$,
and $I(j_1j_2j_3j_4)$ is the radial integral:
\begin{eqnarray}
I(j_1j_2j_3j_4)=N_0^{-1}\int_0^\infty  F_0(r)
u_{j_1}(r)u_{j_2}(r)u_{j_3}(r)u_{j_4}(r)\frac{dr}{r^2},
\end{eqnarray}

where ${\LARGE u(r)}$ is the radial part of the HF single-particle wavefunction.
As it is shown in \cite{gsv98,ssvg02} the radial integrals can be calculated
accurately by choosing a large enough cutoff radius $R$
and using a $N$-point integration Gauss formula with abscissas and weights
${r_k,w_k}$.
\begin{eqnarray}
I(j_1j_2j_3j_4)\simeq N_0^{-1}\frac{R}{2}\sum_{k=1}^{N}
{\frac{w_kF_0(r_k)}{r_k^2}}
u_{j_1}(r_k)u_{j_2}(r_k)u_{j_3}(r_k)u_{j_4}(r_k)
\end{eqnarray}

So we employ the hamiltonian
including an average nuclear HF field,
pairing interactions, the isoscalar and
isovector particle--hole residual forces
in the finite rank separable form \cite{ssvg02}.
This hamiltonian has the same form as the QPM hamiltonian
with $N $ separable terms \cite{solo,sol89},
but in contrast to the QPM all parameters
of this hamiltonian are expressed through parameters of the
Skyrme forces.

In what follows we work in the quasiparticle  representation defined by
the canonical Bogoliubov transformation:
\begin{equation}
a_{jm}^{+}\,=\,u_j\alpha _{jm}^{+}\,+\,(-1)^{j-m}v_j\alpha _{j-m}.
\label{B}
\end{equation}
The single-particle states are specified by the
quantum numbers $(jm)$
The quasiparticle energies,
the chemical potentials, the energy gap and the coefficients
$u$,$v$ of the  Bogoliubov transformations
(\ref{B}) are determined
from  the BCS equations.

We introduce the phonon creation operators
\begin{equation}
Q_{\lambda \mu i}^{+}\,=\,\frac 12\sum_{jj^{^{\prime }}}\left( X
_{jj^{^{\prime }}}^{\lambda i}\,A^{+}(jj^{^{\prime }};\lambda \mu
)-(-1)^{\lambda -\mu }Y _{jj^{^{\prime }}}^{\lambda i}\,A(jj^{^{\prime
}};\lambda -\mu )\right)
\end{equation}

where
\begin{equation}
A^{+}(jj^{^{\prime }};\lambda \mu )\,=\,\sum_{mm^{^{\prime }}}\langle
jmj^{^{\prime }}m^{^{\prime }}\mid \lambda \mu \rangle \alpha
_{jm}^{+}\alpha _{j^{^{\prime }}m^{^{\prime }}}^{+}.
\end{equation}

The index $\lambda $ denotes total angular momentum and $\mu $ is
its z-projection in the laboratory system.
One assumes that the quasiparticle RPA (QRPA) ground state  is the phonon vacuum
$\mid 0\rangle $, i.e. $Q_{\lambda \mu i}\mid 0\rangle\,=0$.
We define the excited states for this approximation by
$Q_{\lambda\mu i}^{+}\mid0\rangle$.

Making use of the linearized equation-of-motion approach \cite{R70}
one can derive the QRPA equations \cite{Schuck,solo}:

\begin{equation}
\label{eq14}
\left(
\begin{tabular}{ll}
${\cal A}$ & ${\cal B}$ \\
${- \cal B}$ & ${- \cal A}$
\end{tabular}
\right) \left(
\begin{tabular}{l}
$ X $ \\
$ Y $
\end{tabular}
\right) =w \left(
\begin{tabular}{l}
$ X $ \\
$ Y $
\end{tabular}
\right).
\end{equation}

In QRPA problems there appear two types of interaction matrix elements,
the ${\cal A}^{(\lambda)}_{(j_1j_1^{\prime})(j_2j_2^{\prime})}$
matrix related to forward-going graphs and the
${\cal B}^{(\lambda)}_{(j_1j_1^{\prime})(j_2j_2^{\prime})}$
matrix related to backward-going graphs \cite{ssvg02}.
Solutions of this set of linear equations yield the eigen-energies
and the amplitudes $X,Y$ of the excited states.
A dimension of the matrixes ${\cal A}, {\cal B}$ is a space size of
the two-quasiparticle configurations.

Using the finite rank approximation we need to invert a matrix having
a dimension $4N \times 4N$ independently of the configuration space size.
One can find a prescription how to solve the system (\ref{eq14}) within
our approach in  \cite{gsv98,ssvg02}.
The QRPA equations in the QPM \cite{solo,sol89} have the same form as
the equations derived within our approach\cite{gsv98,ssvg02},
but the single-particle spectrum and
parameters of the p-h residual interaction are
calculated making use of the Skyrme forces.

In this work we use
the standard parametrization SIII \cite{be75} of the Skyrme force.
Spherical symmetry is assumed for
the HF ground states.
It is well
known \cite{KG00,colo2} that the constant gap approximation leads to
an overestimating of occupation probabilities for subshells that are far
from the Fermi level and it is necessary to introduce a cut-off in the
single-particle space. Above this cut-off subshells don't participate in
the pairing effect. In our calculations we choose the BCS subspace
to include all subshells lying below 5 MeV.
The pairing constants are fixed to reproduce the odd-even mass
difference of neighboring nuclei.
In order to perform RPA calculations, the single-particle continuum is
discretized \cite{BG77} by diagonalizing the HF hamiltonian on a basis
of twelve
harmonic oscillator shells and cutting off the single-particle spectra at the
energy of 190 MeV. This is sufficient to exhaust practically all the
energy-weighted sum rule.

Our investigations \cite{ssvg02} enable us to conclude that $N$=45 is enough for
multipolarities $\lambda \le 3 $ in nuclei with $A\le 208$.
Increasing $N$, for example, up to $N$=60 in $^{208}$Pb does not change results
for energies
and transition probabilities practically.
  Our calculations show
that, for the normal parity states one can neglect the spin-multipole
interactions  as a rule and this reduces by a factor 2
the total matrix dimension.
For heavy nuclei our approach gives a large gain in comparison
with an exact diagonalization \cite{ssvg02}.

To take into account the mixing of the configurations
in the simplest case one can write the wave functions of excited
states as:

\begin{equation}
\Psi _\nu (\lambda \mu )=\{\sum_iR_i(\lambda \nu )Q_{\lambda \mu
i}^{+}+\sum_{\lambda _1i_1\lambda _2i_2}P_{\lambda _1i_1}^{\lambda
_2i_2}(\lambda \nu )\left[ Q_{\lambda _1\mu _1i_1}^{+}Q_{\lambda _2\mu
_2i_2}^{+}\right] _{\lambda \mu }\}|0\rangle  \label{2ph}
\end{equation}

with the normalization condition:
\begin{eqnarray}
\langle \Psi _\nu (JM) \mid \Psi _\nu (JM)\rangle = \sum\limits_iR_i^2(J\nu
)+
2\sum_{\lambda _1i_1 \lambda _2i_2} (P_{\lambda _2i_2}^{\lambda _1i_1}(J\nu
))^2=1
\end{eqnarray}

The matrix element coupling one- and two-phonon configurations is:

\begin{equation}
\langle Q_{Ji }|H|\left[ Q_{\lambda _1i_1}^{+}Q_{\lambda
_2i_2}^{+}\right] _J\rangle =U_{\lambda _2i_2}^{\lambda _1i_1}(Ji)
\end{equation}
 $U_{\lambda _2i_2}^{\lambda _1i_1}(Ji)$ is some combination of the
 geometrical factors and the QRPA phonon amplitudes \cite{solo,vs83}.

The energies of the states $\Psi_\nu (\lambda \mu )$ are solutions of
the following equation \cite{solo}:
\begin{equation}
F(E_\nu )\equiv det\left| (\omega _{\lambda i}-E_\nu )\delta _{ii^{\prime
}}-\frac 12\sum_{\lambda _1i_1,\lambda _2i_2}\frac{U_{\lambda
_2i_2}^{\lambda _1i_1}(\lambda i)U_{\lambda _2i_2}^{\lambda _1i_1}(\lambda
i^{\prime })}{\omega
_{\lambda _1i_1}+\omega _{\lambda _2i_2}
-E_\nu }\right| =0
\label{2-33}
\end{equation}

The rank of the determinant equals the number of the
one-phonon configurations included in the first term of the wave function
$\Psi _\nu (\lambda \mu )$.

It is worth to point out that after
solving the RPA problem with a separable interaction, to take into
account
the coupling with two-phonon configurations demands to diagonalize a
matrix having a size that does not exceed 40 even for the giant resonance
calculations in heavy nuclei whereas one would need to diagonalize a matrix
with a dimension of the order of a few thousands at least for a
non-separable case.

\section{RESULTS OF CALCULATIONS}
As an example we consider  the $2^+_1$, $3^-_1$ state energies and transition
probabilities $B(E\lambda)$ in $^{112}$Sn.
The experimental data \cite{Ram01,Sp02} and the results of our calculations
within the QRPA (the second line) and with taking into account the two-phonon terms
(the third line)
are shown in Table 1. In our calculations the two-phonon terms of the wave
function (\ref{2ph}) include phonons with multipolarities $\lambda = 2,3,4,5$.
One can see that there is
 some overestimate of the energies  and transition probabilities for the QRPA calculations.
The inclusion of the two-phonon configurations results in a reduction of the enrgies and
transition probabilities for the $2^+_1, 3^-_1$ states in $^{112}$Sn.
Generally there is a reasonable agreement between theory
and experiment. The study of an influence of a choice for the Skyrme
forces parameters on properties of the low-lying states within our approach
and calculations for other nuclei are in progress now.

\begin{table}[htb]
\caption{Energies and B(E$\lambda $)-values for up-transitions to the first
2$^+$, 3$^-$ states in $^{112}$Sn.}
\label{tab2}
\begin{tabular}{ccccc}
\hline\\
State & \multicolumn{2}{c}{2$_1^{+}$} &
\multicolumn{2}{c}{3$_1^{-}$} \\\hline\\
& Energy & B(E2) & Energy & B(E3) \\
& (MeV) & (e$^2$fm$^4$) & (MeV) & (e$^2$fm$^6$)
\\
\hline\\
EXP. & 1.26 & 2400$\pm$140 & 2.36& 87000 $\pm$12000 \\
\hline\\
QRPA & 1.49& 2600  & 2.73 & 97000 \\
\hline\\
2PH& 0.90 & 2200 & 1.90 & 72000 \\
 \hline
\end{tabular}
\end{table}

\section{CONCLUSION}

A finite rank separable approximation for the QRPA
calculations with Skyrme interactions that was proposed in
our previous work is extended to take into account the coupling between one-
and two-phonon terms in the wave functions of excited states.

It is shown that the suggested approach enables
one to reduce remarkably the dimensions of the matrices that
must be inverted to perform structure calculations in very large
configuration spaces.

As an illustration of the method we have calculated the
energies and transition probabilities of the  $2^+_1$ and $3^-_1$
states in $^{112}$Sn.

They are in a reasonable agreement with experimental data.
A systematical study of an influence of the two-phonon terms on properties of the
low lying states is in progress now.

\section{ACKNOWLEDGMENTS}

A.P.S. and V.V.V. thank the hospitality of IPN-Orsay where a part
of this work was done.
This work is partly supported
by INTAS Fellowship grant for Young Scientists
(Fellowship Reference N YSF 2002-70),
by IN2P3-JINR agreement
and by the Bulgarian Science Foundation (contract Ph-801).

\end{document}